\documentclass[11pt,a4paper]{article}
\usepackage{jheppub}
\usepackage{axodraw}
\usepackage{bm}

\input pix.sty

\newcommand{\sumint}[1]{\mbox{$\sum$}\!\!\!\!\!\!\!\int_{#1}}
\renewcommand{\eq}{Eq.~}
\renewcommand{\nr}[1]{(\ref{#1})}
\renewcommand{\eqs}{Eqs.~}

\newcommand{\dd}{\mathrm{d}}
\newcommand{\tinymsbar}{{\overline{\mbox{\tiny\rm{MS}}}}}

\newcommand{\Nc}{N_{\rm c}}

\newcommand{\gB}{g_\rmii{B}}

\def\lsi{\raise0.3ex\hbox{$<$\kern-0.75em\raise-1.1ex\hbox{$\sim$}}}
\def\gsi{\raise0.3ex\hbox{$>$\kern-0.75em\raise-1.1ex\hbox{$\sim$}}}

\newcommand{\nB}{n_\rmii{B}}
\newcommand{\rmii}[1]{{\mbox{\tiny\rm{#1}}}}

\newcommand{\Tint}[1]{{\hbox{$\sum$}\!\!\!\!\!\!\!\int\,}_{\!\!\!\!\raise-0.9ex\hbox{$\scriptstyle{#1}$}}}
\newcommand{\Tinti}[1]{{{\Sigma}\!\!\!\!\raise0.3ex\hbox{$\int$}_\rmii{${#1}$}}}
\renewcommand{\Tint}[1]{\sumint{#1}}

\newcommand{\bi}{\begin{itemize}}
\newcommand{\ei}{\end{itemize}}

\newcommand{\hide}[1]{ }

\newcommand{\Jt}[2]{\mathcal{J}_\rmi{#1}^\rmi{#2}}
\newcommand{\It}[2]{\mathcal{I}_\rmi{#1}^\rmi{#2}}

\def\TAsc(#1,#2)(#3,#4,#5)%
{\SetWidth{2.0}\CArc(#1,#2)(#3,#4,#5)\SetWidth{1.0}}
\def\Lwidth{3}

\def\TAgl(#1,#2)(#3,#4,#5){\SetWidth{2.0}\PhotonArc(#1,#2)(#3,#4,#5){\Lwidth}%
{6.283 #3 mul 360 div #4 #5 sub #4 #5 sub mul sqrt mul Tdensity mul}%
\SetWidth{1.0}}
\def\TLgl(#1,#2)(#3,#4){\SetWidth{2.0}\Photon(#1,#2)(#3,#4){\Lwidth}
{#1 #3 sub #1 #3 sub mul #2 #4 sub #2 #4 sub mul add sqrt Tdensity mul}%
\SetWidth{1.0}}

\renewcommand{\picc}[1]{\;\parbox[c]{60pt}{\begin{picture}(60,30)(0,0)
\SetWidth{1.0}\SetScale{1.3} #1 \end{picture}}\; }
\def\Lwidth{1.3}

\newcommand{\picu}[1]{\;\parbox[c]{60pt}{\begin{picture}(60,30)(0,0)
\SetWidth{1.0}\SetScale{1.0} #1 \end{picture}}\; }
\def\EleA{\picu{%
 \Agl(30,5)(22.3,27,153)%
 \Agl(30,25)(22.3,207,333)%
 \COval(10,15)(2,2)(0){Black}{Black}%
 \COval(50,15)(2,2)(0){Black}{Black}%
}}
\def\EleB{\picu{%
 \Agl(30,5)(22.3,27,153)%
 \Agl(30,25)(22.3,207,333)%
 \COval(10,15)(2,2)(0){Black}{Black}%
 \COval(50,15)(2,2)(0){Black}{Black}%
 \Agl(58,15)(8,0,360)%
}}
\def\EleC{\picu{%
 \Agl(30,5)(22.3,27,153)%
 \Agl(30,25)(22.3,207,333)%
 \COval(10,15)(2,2)(0){Black}{Black}%
 \COval(50,15)(2,2)(0){Black}{Black}%
 \Lgl(10,15)(50,15)%
}}
\def\EleD{\picu{%
 \Agl(30,5)(22.3,90,153)%
 \Agl(30,25)(22.3,207,333)%
 \COval(10,15)(2,2)(0){Black}{Black}%
 \COval(50,15)(2,2)(0){Black}{Black}%
 \Agl(43,27)(12,180,300)%
 \Agl(38,16)(12,0,120)%
}}
\def\EleE{\picu{%
 \Agl(30,5)(22.3,27,153)%
 \Agl(30,25)(22.3,207,333)%
 \COval(10,15)(2,2)(0){Black}{Black}%
 \COval(50,15)(2,2)(0){Black}{Black}%
 \GCirc(30,27.3){4}{0.5}
}}
\def\EleF{\picu{%
 \Agl(20,10)(11.15,27,153)%
 \Agl(20,20)(11.15,207,333)%
 \Agl(40,10)(11.15,27,153)%
 \Agl(40,20)(11.15,207,333)%
 \COval(10,15)(2,2)(0){Black}{Black}%
 \COval(50,15)(2,2)(0){Black}{Black}%
}}
\def\EleG{\picu{%
 \Agl(30,5)(22.3,27,153)%
 \Agl(30,25)(22.3,207,333)%
 \COval(10,15)(2,2)(0){Black}{Black}%
 \COval(50,15)(2,2)(0){Black}{Black}%
 \Lgl(30,2.7)(30,27.3)%
}}

\title{The ultraviolet limit and sum rule for the shear correlator in hot Yang-Mills theory}

\author[1]{Y.~Schr\"oder,}
\author[2]{M.~Veps\"al\"ainen,}
\author[1]{A.~Vuorinen}
\author[1]{and Y.~Zhu}

\affiliation[1]{Faculty of Physics, University of Bielefeld, D-33501 Bielefeld, Germany}
\affiliation[2]{Department of Physics, P.O.Box 64, FI-00014 University of Helsinki, Finland}

\emailAdd{yorks@physik.uni-bielefeld.de}
\emailAdd{mikko.vepsalainen@helsinki.fi}
\emailAdd{vuorinen@physik.uni-bielefeld.de}
\emailAdd{yzhu@physik.uni-bielefeld.de}

\abstract{We determine a next-to-leading order result for the correlator of the shear stress operator in high-temperature Yang-Mills theory. The computation is performed via an ultraviolet expansion, valid in the limit of small distances or large momenta, and the result is used for writing operator product expansions for the Euclidean momentum and coordinate space correlators as well as for the Minkowskian spectral density. In addition, our results enable us to confirm and refine a shear sum rule originally derived by Romatschke, Son and Meyer.}


\preprint{BI-TP 2011/39}

\begin{document}

\maketitle

\section{Introduction}

The shear viscosity is one of the most important parameters characterizing the transport properties of the quark gluon plasma. Despite its demonstrated effect on the hydrodynamic expansion of the fireball produced in a heavy ion collision (see e.g.~Refs.~\cite{Romatschke:2009im,Shen:2011zc} and references therein), its first principles determination has proved notoriously challenging. As lattice QCD is limited to the Euclidean formulation of the theory and weak coupling expansions of transport constants are tedious to perform and display slow convergence \cite{Arnold:2003zc}, much attention has lately shifted to gauge gravity calculations. This has in particular led to the conjecture of the quark gluon plasma being a nearly `ideal' fluid, with a shear viscosity to entropy ratio close to the famous limit of $\eta/s\geq 1/(4\pi)$ \cite{Kovtun:2004de}. It would, nevertheless, clearly be of considerable value to have a quantitative estimate for the quantity starting from truly first principles, i.e.~the QCD Lagrangian.

At present, the arguably most promising approach to determine the shear and bulk viscosities in QCD utilizes lattice measurements of Euclidean energy momentum tensor correlators $G_n(x)$, which so far have been carried out only in pure Yang-Mills theory \cite{Meyer:2007ic,Meyer:2007dy,Huebner:2008as}. Specializing to the shear channel, $n=\eta$, one uses this data in an attempt to invert the relation
\begin{equation}
 G_\eta(\hat \tau) =
 \int_0^\infty
 \frac{{\rm d}\omega}{\pi} \rho_\eta(\omega)
 \frac{\cosh\big[\! \left(\frac{1}{2} - \hat\tau\right)\beta\omega\big]}
 {\sinh\frac{\beta \omega}{2}}\, ,\quad \quad 0< \hat{\tau}<1\, ,  \la{int_rel}
\end{equation}
to gather information on the spectral density, from the intercept of which the corresponding viscosity is obtained, $\eta = \lim_{\omega\to 0} \fr{\rho_\eta(\omega)}{\omega}$. To this end, Refs.~\cite{Romatschke:2009ng,Meyer:2010gu} derived a convenient sum rule (cf.~Eq.~(1) in each reference and Eq.~(\ref{rule}) below), which can be used to constrain the form of $\rho_\eta(\omega)$. This rule contains a contact term, which is obtainable from the UV-limit of a finite-temperature correlator that is most conveniently evaluated in the continuum, using perturbation theory and dimensional regularization. This calculation necessitates in particular the determination of the corresponding limit of the shear operator ($T_{12}$) correlation function.

In addition to sum rules, correlation functions of the energy momentum tensor are interesting in their own right. In Refs.~\cite{Laine:2010tc,Laine:2010fe,Laine:2011xm}, a systematic program was initiated for the next-to-leading order (NLO) perturbative evaluation of the Euclidean correlators and (Minkowskian) spectral densities corresponding to various components of the energy momentum tensor in SU($N$) Yang-Mills theory. So far, the calculations have been limited to the bulk channel (the scalar and pseudoscalar operators), in which they have been applied first to the UV limit of the Euclidean correlator \cite{Laine:2010fe}, and later to the full time-averaged spatial correlator \cite{Laine:2010tc} and the spectral density \cite{Laine:2011xm}. Direct comparisons with lattice data and gauge/gravity results (see e.g.~Refs.~\cite{Iqbal:2009xz,Springer:2010mw}) are possible in all of these cases, and are hoped to shed at least some light on the question, whether quark gluon plasma can be characterized as weakly or strongly coupled at temperatures slightly above the deconfinement one.

In the present paper, our aim is to determine the shear correlator, defined in Eq.~(\ref{corrdef}) below and referred to as the \textit{tensor} channel correlator in some references, up to NLO in simultaneous weak coupling and UV expansions. The result of this calculation will enable us to verify the sum rule of Refs.~\cite{Romatschke:2009ng,Meyer:2010gu}, in which we identify an additional order $g^4$ contribution that unfortunately is beyond the accuracy of our computation. We also evaluate the short distance limit of the equal time correlator in configuration space and the large frequency limit of the corresponding spectral function. The latter result will serve as a useful test for the prediction of Ref.~\cite{CaronHuot:2009ns}, according to which the leading temperature-dependent corrections to the vacuum spectral density in the shear channel should vanish in the absence of fermions. Relevant work on the shear correlator has in addition been carried out in holographic models, cf.~e.g.~Refs.~\cite{Springer:2010mf,Kajantie:2010nx,Kajantie:2011nx}.

The outline of the paper is as follows. In Section 2, we explain our setup and define the quantities we set out to evaluate. Section 3 then contains a brief outline of the calculations, while in Section 4 we collect our results for the UV limit of the Euclidean correlator and apply them to the determination of the equal time spatial correlator as well as to the spectral density and the shear sum rule. In Section 5, we finally draw our conclusions. Some calculational details, in particular the detailed evaluation of one of the new sum-integrals encountered in the calculation, are left to the appendices.

\section{Setup and definitions}

Linear response theory describes the small departures from thermal equilibrium that arise when a physical system is subjected to an external perturbation. In particular, it relates changes in physical observables to the retarded correlators of the corresponding operators. When the energy and momenta are small, the behavior of the correlation functions can be described using hydrodynamics, which can be viewed as a low energy effective theory for the underlying thermal field theory. The matching constants between these two theories are known as transport coefficients, and can be computed from the zero-momentum limits of retarded correlation functions using so-called Kubo formulae.

The operator relevant for the computation of the bulk and shear viscosities is the energy-momentum tensor $T_{\mu\nu}$. The general tensorial structure of its correlators has been worked out in detail in Ref.~\cite{Kovtun:2005ev}, assuming symmetry in the indices, rotational invariance in $D-1$ spatial dimensions,\footnote{We work in $D=4-2\epsilon$ Euclidean space-time dimensions.} and transversality with respect to the external momentum $P$ (current conservation). However, depending on the definition, the correlator we compute may not be fully transversal. As pointed out by e.g.~Refs.~\cite{Kovtun:2005ev,Romatschke:2009ng}, $P_\mu \:\langle T_{\mu\nu}(P) T_{\alpha\beta}(-P)\rangle$ may differ from zero by terms polynomial in $P$, which translate to contact terms in coordinate space. This is precisely what happens for the correlator conventionally computed in perturbation theory and on the lattice, and it is for this reason that we do not apply the general decomposition of Ref.~\cite{Kovtun:2005ev} here. Instead, we project out the desired correlators without any reference to transversality.

In this paper, we work with pure Yang-Mills theory at finite temperature. The dimensionally regularized Euclidean action is written as
\begin{equation}
 S_\mathrm{E} = \int_{0}^{\beta} \! \dd \tau \int \! {\rm d}^{3-2\epsilon}\vec{x}
 \, \frac{1}{4} F^a_{\mu\nu} F^a_{\mu\nu} \,,
\end{equation}
where we have defined
$ F^a_{\mu\nu} = (2i/\gB) \tr \{ T^a [D_\mu, D_\nu] \} =
 \partial_\mu A^a_\nu - \partial_\nu A^a_\mu + \gB f^{abc} A^b_\mu A^c_\nu $ and
$ D_\mu = \partial_\mu - i \gB A^a_\mu T^a $.
The energy-momentum tensor of the theory reads correspondingly
\begin{equation}\la{eq:T}
 T_{\mu\nu} = \frac{1}{4} \delta_{\mu\nu} F^a_{\alpha\beta} F^a_{\alpha\beta} -F^a_{\mu\alpha} F^a_{\nu\alpha} \, ,
\end{equation}
and the correlators of this operator that we are interested in are defined through
\begin{equation}
 G_{\mu\nu,\alpha\beta}(x)  \equiv \langle T_{\mu\nu}(x)\: T_{\alpha\beta}(0)\rangle_c\, .
\label{eq:corr_def_coord}
\end{equation}
Here, the symbol $\langle\ldots\rangle_c$ stands for the connected part of the thermal correlation function. The corresponding momentum space expression reads
\begin{equation}
 \tilde G_{\mu\nu,\alpha\beta}(P) \equiv \int_x e^{-iP\cdot x} G_{\mu\nu,\alpha\beta}(x) \, ,
\label{eq:corr_def_mom}
\end{equation}
where the Fourier transform should be taken in $D$ dimensions.

To evaluate the shear channel correlator, $G_{12,12}(x)$, we introduce the projection operator
\begin{equation} \label{eq:xproj}
    X_{\mu\nu,\alpha\beta} \equiv P_{\mu\nu}^{T}P_{\alpha\beta}^{T}
	-\frac{D-2}{2}(P_{\mu\alpha}^{T}P_{\nu\beta}^{T}+P_{\mu\beta}^{T}P_{\nu\alpha}^{T}) \,,
\end{equation}
where $P_{\mu\nu}^{T}$ is a symmetric projector orthogonal to $P$ as well as the four vector $U=(1,\mathbf{0})$,
\begin{align}
P_{\mu\nu}^{T} &\equiv \delta_{\mu\nu}-\frac{P_\mu P_\nu
+P^2 U_\mu U_\nu-(UP)\(U_\mu P_\nu +U_\nu P_\mu\)}{P^2-(UP)^2}\;,\\
\Rightarrow &\quad
P_\mu P_{\mu\nu}^{T} = 0 = U_\mu P_{\mu\nu}^{T}\;,\quad
P_{\mu\alpha}^{T}P_{\alpha\nu}^{T}=P_{\mu\nu}^{T}\;,\quad
P_{\mu\mu}^{T}=D-2\;.
\la{eq:P}
\end{align}
In components, $P_{00}^T=0=P_{0i}^T$,
$P_{ij}^T = \delta_{ij}-\frac{p_i p_j}{\mathbf{p}^2}$.
Applying $X_{\mu\nu,\alpha\beta}$ to the correlator in
Eq.~(\ref{eq:corr_def_mom}) and choosing the spatial momentum $\mathbf{p}$
along the $x_{D-1}$-direction, we obtain
\begin{equation}
    X_{\mu\nu,\alpha\beta}\: \tilde G_{\mu\nu,\alpha\beta}(P) = -D(D-2)(D-3)\: \tilde G_{12,12}(P) \,,
\end{equation}
where we have exploited rotation invariance in $D-2$ dimensions. As suggested by this result, we define the Euclidean correlation function that we will evaluate below by
\begin{equation}
 G_\eta(x) \equiv 2 c_\eta^2 X_{\mu\nu,\alpha\beta}(x) \, \langle T_{\mu\nu}(x)\: T_{\alpha\beta}(0)\rangle_c \,, \label{eq:g_eta_p}
\end{equation}
where $X_{\mu\nu,\alpha\beta}(x) $ denotes the corresponding projector in coordinate space and we have (mainly for historical reasons; see the notation of Ref.~\cite{Laine:2010tc}) introduced a coefficient $c_\eta$, whose value
can be chosen at will. When we set $D=4$ and choose the spatial separation along the $x_3$-direction, the correlator reduces to the simple expression
\begin{equation}
 G_\eta(x)   = -16 c_\eta^2  \: \langle T_{12}(x)\: T_{12}(0)\rangle_c\, . \label{corrdef}
\end{equation}

\section{Details of the calculation}

To evaluate the correlation function of Eq.~(\ref{eq:g_eta_p}) in momentum space, we use \eqs\nr{eq:T}-\nr{eq:P} to obtain
\ba\la{eq:gEtaP}
\tilde G_\eta(P)=2c_\eta^2\, X_{\mu\nu,\alpha\beta} \int_x e^{-iP\cdot x}
\,\langle F_{\mu\rho}^a(x)F_{\nu\rho}^a(x)\,F_{\alpha\sigma}^b(0)F_{\beta\sigma}^b(0)\rangle_c\;.
\ea
In expanding this expression to NLO in perturbation theory, we follow standard procedures: We start by generating the relevant one- and two-loop graphs, displayed in Fig.~\ref{fig:graphs}, with QGRAF \cite{Nogueira:1991ex}, and next insert into them the Feynman rules and carry out the Lorentz and color algebra. Symmetrization and decoupling of momenta, i.e.~removing as much of the numerator structure as possible, leaves now a considerably larger set of sum-integrals to be evaluated than in the bulk channel case. Keeping $D$ unspecified for the moment, and using the gluon propagator in the covariant gauge with gauge parameter $\xi$, we obtain
\ba
\frac{\tilde G_\eta(P)}{4 d_A c_\eta^2\Lambda^{2\epsilon}} &=&
\fr{D(D-2)(D-3)}{8}\(2\Jt{10}{0}-\Jt{11}{0}\) -(D-2)(D-3)\Jt{11}{2}
+D(D-3)\(\Jt{10}{1}-\Jt{11}{1}\) \nn
&+& \gB^2 \Nc \Biggl\{
\fr{D(D-2)(D-3)}{4}\bigg(2 \It{11000}{000} + 4 \It{11001}{100} + 2
\It{11001}{010} + 4 \It{12001}{100} + 12 \It{12001}{010}
\nn&+&2 \It{11100}{000}
+ \It{21100}{100} + 4 \It{21100}{010}
 - 3 \It{10101}{100} - 4 \It{11101}{000} - 4 \It{11101}{100} + \It{12101}{010}
 - 4 \It{111-11}{100} \nn
 &-&4 \It{121-11}{100} - 2 \It{121-21}{010} + \It{11111}{000}\bigg)\nonumber \\
&+&\fr{D(D-2)}{2}\bigg(-\It{21100}{200} - 2 \It{21100}{020} + 4 \It{21100}{110}
+2 \It{21100}{011} + 2 \It{12101}{200} + \It{12101}{020} \nn
&-& 4 \It{12101}{110} - 2 \It{12101}{101}\bigg)
-\fr{(D-2)^2(D-3)}{4}\bigg(D \It{12000}{000} - D \It{12001}{000} - 8
\It{12001}{020}\bigg)
\nn&+&D(D-3)\bigg(-4\It{11101}{010}+2\It{11111}{100}+\It{11111}{001}\bigg)
-D(D-6)\bigg(\It{11111}{110}+2\It{11111}{101}\bigg)
\nn&+&\fr{12-16D+3D^2}{2}\bigg(2\It{11111}{200}+\It{11111}{002}\bigg)
+\fr{D(D-3)(3D-10)}{4}\It{11100}{100} \Biggr\}\,, \label{Gmasters}
\ea
where the one- and two-loop sum-integrals $\Jt{}{}$ and $\It{}{}$ are given by
\ba
\la{eq:defJ}
\mathcal{J}_{ab}^{c} & \equiv& \Tint{Q}
\frac{[P_T(Q)]^c[P^2]^{a+b-c}}
{[Q^2]^a[(Q-P)^2]^b} \,, \\
\la{eq:defI}
\mathcal{I}_{abcde}^{fgh} & \equiv& \Tint{Q,R}
\frac{[P_T(Q)]^f[P_T(R)]^g[P_T(Q-R)]^h[P^2]^{a+b+c+d+e-f-g-h-2}}
{[Q^2]^a[R^2]^b[(Q-R)^2]^c[(Q-P)^2]^d[(R-P)^2]^e} \,.
\ea
These integrals are defined using dimensional regularization in $D=4-2\epsilon$ dimensions, and we employ the shorthands $P_T(Q) \equiv Q_\mu Q_\nu P^T_{\mu\nu}(P) = \mathbf{q}^2-(\mathbf{q\cdot \hat p})^2$ and
$\Tinti{Q} \equiv T \sum_{q_0} \int_\vec{q}$ with $\int_\vec{q}\equiv\int\!{\rm  d}^{D\!-\!1}\vec{q}/(2\pi)^{D\!-\!1}$. Note also that the gauge parameter $\xi$ has cancelled in the sum of the diagrams, serving as a nice check of the computation.

%
\begin{figure}[t]
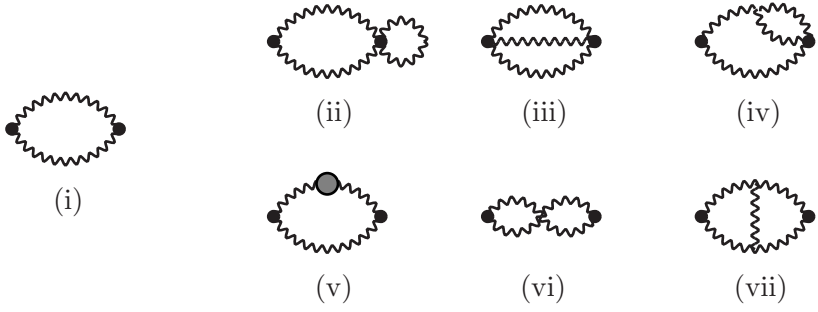


\hspace*{1.5cm}%
\begin{minipage}[c]{3cm}
\begin{eqnarray*}
&&
 \hspace*{-1cm}
 \EleA
\\[1mm]
&&
 \hspace*{0.0cm}
 \mbox{(i)}
\end{eqnarray*}
\end{minipage}%
\begin{minipage}[c]{10cm}
\begin{eqnarray*}
&&
 \hspace*{-1cm}
 \EleB \quad\;
 \EleC \quad\;
 \EleD \quad\;
\\[1mm]
&&
 \hspace*{0.0cm}
 \mbox{(ii)} \hspace*{2.2cm}
 \mbox{(iii)} \hspace*{2.2cm}
 \mbox{(iv)}
\\[5mm]
&&
 \hspace*{-1cm}
 \EleE \quad\;
 \EleF \quad\;
 \EleG \quad
\\[1mm]
&&
 \hspace*{0.0cm}
 \mbox{(v)} \hspace*{2.2cm}
 \mbox{(vi)} \hspace*{2.2cm}
 \mbox{(vii)}
\end{eqnarray*}
\end{minipage}
\caption[a]{\small
The LO and NLO Feynman graphs contributing to the correlators of the energy momentum tensor.}
\la{fig:graphs}
\end{figure}
%

The evaluation of the specific sum-integrals present in \eq\nr{Gmasters} proceeds with methods closely analogous to those developed and explained in some length in Ref.~\cite{Laine:2010tc}. One starts by performing the Matsubara sums, thus dividing the result into three parts proportional to zero, one and two Bose distribution functions. The next steps involve then an expansion of the result in positive powers of the momenta exponentially cut off by thermal distribution functions, and the subsequent evaluation of the remaining integrals. To highlight the differences with respect to the bulk channel calculation, we will in Appendix \ref{app:example} go through the evaluation of one of the new sum-integrals, $\It{12101}{101}$, in detail. For the others, we display only the result obtained after summing all of the terms in Eq.~(\ref{Gmasters}) together, cf.~Eq.~(\ref{res1}) below. This is in line with our general strategy of automatizing as much of the calculation as possible and all the time dealing with the full expression for the correlator rather than individual sum-integrals.

\section{Results}

\subsection{The Euclidean momentum space correlator}

Following the steps outlined in the previous section, we arrive at a result for the sum of all integrals in Eq.~(\ref{Gmasters}). This leads to the Euclidean space expression for the shear correlator,
\ba
\frac{\tilde G_\eta(P)}{4 d_A c_\eta^2}&=&\frac{P^4}{(4\pi )^2}\Bigg\{-\frac{2 }{5 } \left(\frac{\bar{\Lambda }}{P}\right)^{2\epsilon }\left(\frac{1}{\epsilon }-\frac{13 }{5 }\right) +\frac{g^2N_c}{(4\pi )^2}\left(\frac{\bar{\Lambda }}{P}\right)^{4\epsilon }\left(\frac{4}{9\epsilon}-\frac{206}{135}+\frac{24 \zeta(3)}{5}\right)\Bigg\} \nonumber
\ea
\ba
&-&\Bigg\{\frac{8}{3} +\frac{8}{P^2}\left(\frac{p^2}{3}-p_n^2\right)
  -\fr{8}{9}\frac{g^2N_c}{(4\pi)^2}\Bigg(22 +\frac{41}{P^2}\left(\frac{p^2}{3}-p_n^2\right)\Bigg)\Bigg\}
  \int_{\mathbf{q}}\nB(q)q\nn
&+&g^2N_c\Bigg\{\fr{20}{3}+\frac{12}{P^2}\left(\frac{p^2}{3}-p_n^2\right)\Bigg\}\int_{\mathbf{q}\,\mathbf{r}}\frac{\nB(q)}{q} \frac{\nB(r)}{r}+{\mathcal O}\(g^4,\fr{T^6}{P^2}\)\,, \label{res1}
\ea
in which $p_n$ is the Matsubara frequency and $p\equiv |\mathbf{p}|$. There are several interesting things to note in this expression. First, the $T=0$ part of the result agrees with the previously known expression obtained in Ref.~\cite{Pivovarov:1999mr}. Furthermore, as expected from the form of the shear operator, necessitating only coupling constant renormalization starting at NNLO, all $1/\epsilon$ divergences have automatically canceled in the $T$-dependent part of the result. This in particular implies that there are no logs of $P^2$ in the thermal terms, which will have important implications on the spectral density, discussed in Section~\ref{4.3}.

We also note that the form of our result for the correlator allows an interpretation of its $T$-dependent part, denoted here by $\Delta\tilde G_\eta(P)$, in terms of an operator product expansion (OPE). Following the notation of Ref.~\cite{Laine:2010tc} (cf.~this reference for additional discussion on OPEs), we obtain
\ba
\frac{\Delta\tilde G_\eta(P)}{4 c_\eta^2}&=&-\Bigg\{1 +\frac{3}{P^2}\left(\frac{p^2}{3}-p_n^2\right)
  -\fr{1}{3}\frac{g^2N_c}{(4\pi)^2}\Bigg(22+\frac{41}{P^2}\left(\frac{p^2}{3}-p_n^2\right)\Bigg)\Bigg\}
  (e+p)(T) \nn
&+&\fr{4}{3g^2b_0}(e-3p)(T)\Big\{1-g^2b_0\ln\zeta_{12}\Big\}+{\mathcal O}\(g^4,\fr{T^6}{P^2}\)\,, \label{res2}
\ea
where $b_0=11N_c/(3(4 \pi)^2)$ originates from the one-loop beta function of the theory, and we have used the thermodynamic identities
\ba
(e+p)(T)&=&\fr{8d_A}{3}\Bigg[\int_{\mathbf{q}}\nB(q)q-\fr{3g^2N_c}{2}\int_{\mathbf{q}\,\mathbf{r}}\frac{\nB(q)}{q} \frac{\nB(r)}{r}\Bigg]\,,\\
(e-3p)(T)&=&2d_A g^4 b_0 N_c \int_{\mathbf{q}\,\mathbf{r}}\frac{\nB(q)}{q} \frac{\nB(r)}{r} \,.
\ea
There is an intriguing subtlety related to the renormalization scale running of the Wilson coefficients in the result. While the one multiplying the operator $(e+p)(T)$ is expected to be scale invariant and thus contain logs of the renormalization scale $\bar{\Lambda}\sim\omega$ at higher orders, the same does not apply to the trace anomaly contribution on the second row of Eq.~(\ref{res2}). As discussed in Ref.~\cite{CaronHuot:2009ns}, in order to obtain a transverse spectral function, this term should be interpreted as a genuine contact term, and the coupling in its Wilson coefficient allowed to run with a renormalization scale $\bar\Lambda$, independent of $\omega$. This in particular implies that the NLO contribution to this function be a mere constant, which we have chosen to parametrize in terms of a coefficient $\zeta_{12}$ (cf.~the constants $\zeta_{\theta}$ and $\zeta_{\chi}$ defined in Ref.~\cite{Laine:2010tc}).


\subsection{The equal time spatial correlator \label{4.2}}

\begin{figure}[t]

\centerline{%
 \epsfysize=8.5cm\epsfbox{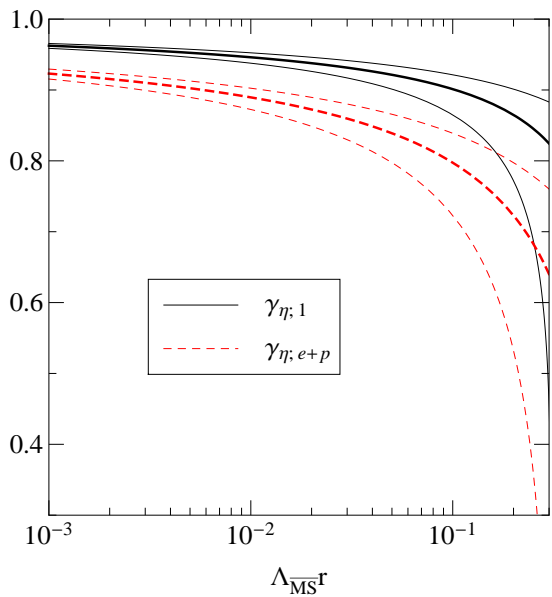}
}

\caption[a]{\small
The behavior of the Wilson coefficients $\gamma_{\eta;1}(r)$ and $\gamma_{\eta;e+p}(r)$ as functions of $\Lambda_{\tinymsbar}\,r$, obtained using a two-loop expression for the gauge coupling $g$. The three curves correspond in each case to the variation of the renormalization scale $\bar{\Lambda}$ around the value $1/r$ by a factor of 2. The central value of the parameter is in accordance with the choices made in Ref.~\cite{Laine:2010tc}; in our case, the method of fastest apparent convergence (FAC) is not applicable.} \label{fig2}

\end{figure}

Next, we use the Euclidean momentum space result to study the corresponding equal time correlator in coordinate space, of which we are able to obtain the two leading terms in a short distance expansion. To this end, we take a four-dimensional ($T=0$) Fourier transform of Eqs.~(\ref{res1})--(\ref{res2}), which using the standard formulae listed in Ref.~\cite{Laine:2010tc} produces
\begin{align}
\frac{G_\eta(r)}{4 c_\eta^2} &= -\fr{24d_A}{5\pi^4
  r^8}\gamma_{\eta;1}(r)+\fr{2(e+p)(T)}{\pi^2
  r^4}\gamma_{\eta;e+p}(r) +\fr{(e-3p)(T)}{\pi^2r^4}\gamma_{\eta;e-3p}(r)
    +{\mathcal O}\(\fr{T^6}{r^2}\) \,,
\end{align}
with
\ba
\gamma_{\eta;1}(r)&=&1-\fr{20}{9}\fr{g^2N_c}{(4\pi)^2}+{\mathcal O}(g^4) \,, \label{gamma1}\\
\gamma_{\eta;e+p}(r)&=&1-\fr{41}{9}\fr{g^2N_c}{(4\pi)^2}+{\mathcal O}(g^4) \,,\\
\gamma_{\eta;e-3p}(r)&=&{\mathcal O}(g^2) \,.\label{gammae3p}
\ea

The behavior of the Wilson coefficients $\gamma_n$ as functions of $\Lambda_{\tinymsbar}\,r$ is plotted in Fig.~\ref{fig2}. Comparing the results to the corresponding bulk channel quantities from Ref.~\cite{Laine:2010tc}, we observe markedly improved convergence. It would be highly interesting to be able to compare the obtained behavior to lattice data, along the lines of Ref.~\cite{Iqbal:2009xz} in the bulk case, as well as later for the full correlator, for which problems with lattice artifacts and discretization would be less pronounced. We are, however, unaware of the existence of such results in the shear channel.

\subsection{The spectral density and the shear sum rule \label{4.3}}

Another natural application of our results is the determination of the UV limit of the shear spectral density $\rho_\eta(\omega)\equiv {\rm Im}\,\tilde{G}_\eta(-i(\omega+i\epsilon),0)$, cf.~the discussion in Ref.~\cite{Laine:2010tc}. Analytically continuing Eq.~(\ref{res2}) to Minkowski space and taking the imaginary part of the result at $\mathbf{p}=0$, we see that to the accuracy of our computation, no $T$-dependent terms survive,  in accordance with the arguments of Ref.~\cite{CaronHuot:2009ns}. This leads to the compact result
\begin{align}
\fr{\rho_\eta(\omega)}{4c_\eta^2 \pi} &= -\fr{2d_A\omega^4}{5(4\pi)^2}\tilde{\gamma}_{\eta;1}(\omega)+(e+p)(T)\tilde{\gamma}_{\eta;e+p}(\omega)+(e-3p)(T)\tilde{\gamma}_{\eta;e-3p}(\omega)
+{\mathcal O}\(\fr{T^6}{\omega^2}\) \,,
\end{align}
where
\ba
\tilde{\gamma}_{\eta;1}(\omega)&=&1-\fr{20}{9}\fr{g^2N_c}{(4\pi)^2}+{\mathcal O}(g^4) \,, \\
\tilde{\gamma}_{\eta;e+p}(\omega)&=&{\mathcal O}(g^4) \label{gammaspe1} \,,\\
\tilde{\gamma}_{\eta;e-3p}(\omega)&=&{\mathcal O}(g^2) \label{gammaspe3}\,.
\ea
The computation of the full NLO spectral density is currently underway; to gain some insight into the features to be expected, we refer the reader to the bulk channel work of Ref.~\cite{Laine:2011xm} as well as to the one-loop HTL result on Ref.~\cite{Aarts:2002cc}.

The spectral density also appears inside a sum rule derived in Refs.~\cite{Romatschke:2009ng,Meyer:2010gu}, which relates its integral to the energy density. In our notation, the rule reads
\ba
-\fr{1}{16\pi c_\eta^2}\int_{-\infty}^{\infty}\fr{{\rm d}\omega}{\omega} \Big\{\rho_\eta(\omega)-\rho_\eta(\omega)\!\mid_{T=0}\Big\}+\lim_{\omega\rightarrow \infty}{\mathcal G}(\omega,T)&=&\fr{2}{3}e(T) \,, \label{rule}
\ea
where the Minkowskian correlator in the contact term, ${\mathcal G}(\omega,T)$, is defined through\footnote{Note that when the corresponding Euclidean results are analytically continued to Minkowski space, we automatically take the real parts of the expressions.}
\ba
{\mathcal G}(\omega,T)&\equiv& -\frac{\Delta\tilde G_\eta(-i\omega,\mathbf{p}=0)}{16 c_\eta^2}-\fr{2}{3}\int {\rm d}^4x\, e^{i\omega x_0}\langle T_{00}(x)T_{00}(0) \rangle_{T}\, , \label{Gdef}
\ea
and the subscript $T$ denotes the finite-temperature part of the function. In Ref.~\cite{Meyer:2010gu}, the entire contact term was argued to be finite and of order ${\mathcal O}(1)\times (e-3p)(T)$.

To inspect the sum rule, we begin by evaluating the contact term. To this end, we have independently checked the result quoted in Ref.~\cite{Meyer:2010gu},
\ba
\int {\rm d}^4x\, e^{i\omega x_0}\langle T_{00}(x)T_{00}(0) \rangle_{T}
    &=& -\fr{3}{4}\Big\{1+{\mathcal O}(g^4)\Big\}(e+p)(T) \label{T00}\\
&-&\fr{1}{2g^2b_0}\bigg\{1-g^2b_0\ln \zeta_{00}\bigg\}(e-3p)(T)+{\mathcal O}\(g^4,\fr{1}{P^2}\) \,, \nonumber
\ea
where $\zeta_{00}$ is yet another unknown three-loop parameter and where the renormalization scales are to be interpreted in the fashion discussed after Eq.~(\ref{res2}). Adding this to the UV limit of the shear correlator according to Eq.~(\ref{Gdef}) and going to the $\omega\rightarrow\infty$ limit, we obtain the simple (and finite) result
\ba
\lim_{\omega\rightarrow \infty}{\mathcal G}(\omega,T)&=&\fr{1}{3}\ln\fr{\zeta_{12}}{\zeta_{00}}\times(e-3p)(T) \,=\, {\mathcal O}(g^4)\, , \label{contact}
\ea
which verifies the claim of Ref.~\cite{Meyer:2010gu} regarding the size of the term. The determination of the coefficients $\zeta_{12}$ and $\zeta_{00}$ is unfortunately beyond the accuracy of our calculation.

Finally, we take a look at the integral containing the spectral function. For it, we may use the relation
\ba
\fr{1}{\pi}\int_{-\infty}^{\infty}\fr{{\rm d}\omega}{\omega} \Big\{\rho_\eta(\omega)-\rho_\eta(\omega)\!\mid_{T=0}\Big\}&=&
\tilde G_\eta(0)-\lim_{\omega\rightarrow \infty} \Delta\tilde G_\eta(-i\omega,\mathbf{p}=0) \,, \label{sum1}
\ea
in which we have used the fact that $\tilde{G}(0)$ vanishes at $T=0$. The first term here involves the zero momentum limit of the (Euclidean) shear correlator and thus reduces to two-loop vacuum type sum-integrals. This calculation is performed in Appendix B and results in
\begin{align} \label{zeromom2}
-\frac{\tilde G_\eta(0)}{16 c_\eta^2} &= -\fr{1}{3g^2b_0}\(e-3p\)(T) + {\mathcal O}(g^4) \,,
\end{align}
which cancels with the leading order part of the contact term on the second row of Eq.~(\ref{res2}). What remains from the sum of Eq.~(\ref{sum1}) is simply the $(e+p)(T)$ part of the UV term, leaving us with
\ba
-\fr{1}{16\pi c_\eta^2}\int_{-\infty}^{\infty}\fr{{\rm d}\omega}{\omega} \Big\{\rho_\eta(\omega)-\rho_\eta(\omega)\!\mid_{T=0}\Big\}&=&
\fr{1}{2}(e+p)(T) +{\mathcal O}(g^4)\nn
&=&\fr{2}{3}e(T)+{\mathcal O}(g^4). \label{specin}
\ea
Here, the unknown ${\mathcal O}(g^4)$ term contains contributions from both the $\ln\zeta_{12}$ term in Eq.~(\ref{res2}) and the unknown three-loop contributions to the result of Eq.~(\ref{zeromom2}).

Looking back to the left hand side of the sum rule of Eq.~(\ref{rule}), we note that the sum of Eqs.~(\ref{contact}) and (\ref{specin}) is in perfect agreement with the right hand side. As explained above, there however appear to be non-vanishing order $g^4$ contributions to the former that our computation has unfortunately been unable to resolve.

\section{Discussion and conclusions}

In the paper at hand, we have perturbatively evaluated the UV limits of the thermal correlators of the shear operator $T_{12}$ and the energy density $T_{00}$ in hot Yang-Mills theory. The calculations were performed to next-to-leading order in both weak coupling and UV expansions, and were subsequently used to determine OPE coefficients for the equal time coordinate space correlator and the spectral density in the shear channel. In addition, we confirmed and refined a sum rule for the shear spectral density, proposed originally in Refs.~\cite{Romatschke:2009ng,Meyer:2010gu}.

Let us briefly summarize our main results. In Section \ref{4.2}, we found that the NLO expressions we obtained for the Wilson coefficients of the coordinate space shear correlator, Eqs.~(\ref{gamma1})--(\ref{gammae3p}), display considerably improved convergence properties in comparison with those obtained in the bulk channel in Ref.~\cite{Laine:2010fe}. For the spectral density, our Eqs.~(\ref{gammaspe1})--(\ref{gammaspe3}) on the other hand verified the claim of Ref.~\cite{CaronHuot:2009ns} that the leading order thermal corrections to the quantity should vanish for $N_\rmi{f}=0$. And finally, in the context of the shear sum rule of Refs.~\cite{Romatschke:2009ng,Meyer:2010gu},
\ba
-\fr{1}{16\pi c_\eta^2}\int_{-\infty}^{\infty}\fr{{\rm d}\omega}{\omega} \Big\{\rho_\eta(\omega)-\rho_\eta(\omega)\!\mid_{T=0}\Big\}&=&\fr{2}{3}e(T)-\lim_{\omega\rightarrow \infty}{\mathcal G}(\omega,T) \,, \label{rule2}
\ea
we independently evaluated all of the terms, finding perfect agreement to the accuracy of our computation. In particular, we found a simple result for the contact term appearing here, cf.~Eq.~(\ref{contact}). The numerical evaluation of the constants $\zeta_{12}$ and $\zeta_{00}$, defined in Eqs.~(\ref{res2}) and (\ref{T00}), would unfortunately require a three-loop determination of the $T_{12}$ and $T_{00}$ correlators, which lies beyond the accuracy of our computation.

There are clearly several directions, to which our present calculations can and should be continued. Most importantly, it would be highly interesting to perturbatively evaluate the \textit{full} shear spectral density, using methods developed in Ref.~\cite{Laine:2011xm}. This would provide important input for the lattice determination of the shear viscosity, as it would enable the analytic subtraction of the leading short distance terms from the lattice data and thus allow for a better modeling of the spectral shape. In addition to this, it would of course be important to consider the effects of fermions, both in lattice simulations and within the corresponding perturbative analysis. In the perturbative setting, the addition of dynamical fermions is in principle straightforward --- though somewhat tedious --- to implement; as a first step, it would be nice to be able to confirm the UV limit of the spectral density as predicted in Ref.~\cite{CaronHuot:2009ns}. And finally, one would naturally like to pursue the evaluation of the coefficients $\zeta_{12}$ and $\zeta_{00}$, which are the only parts missing from the determination of the contact term in the shear sum rule. We plan to address these issues in future work.

\section*{Acknowledgments}
We are indebted to Mikko Laine for his valuable advice and comments, as well as to Keijo Kajantie and Harvey B.~Meyer for useful discussions. Y.S.~is supported by the Heisenberg program of the Deutsche Forschungsgemeinschaft (DFG), contract no.~SCHR 993/1, M.V.~by the Academy of Finland, contract no.~128792, A.V.~by the Sofja Kovalevskaja program of the Alexander von Humboldt foundation, and Y.Z.~by the DFG International Graduate School \textit{Quantum Fields and Strongly Interacting Matter}.

\appendix

\section{The two-loop sum-integral $\It{12101}{101}$} \label{app:example}

In this Appendix, we present the detailed evaluation of the two-loop
sum-integral
\ba
\It{12101}{101} & \equiv &
 \Tint{Q,R} \frac{P^2}{Q^2R^4(Q-R)^2(R-P)^2}P_T(Q)P_T(Q-R)
 \,,
\ea
which serves as an illustrative example of the subtleties related to having extra numerator structures in the integrand. Carrying out the Matsubara sums in this expression gives after some straightforward algebra
\ba
\label{Ih7c}
\It{12101}{101} & = & \int_{Q,R} \frac{P^2}{Q^2R^4(Q-R)^2(R-P)^2}P_T(Q)P_T(Q-R)\nn
&+&\int_{\mathbf{q}}\frac{\nB(q)}{q}\,\Bigg[\int_{R}
\bigg( \frac{P^2\,f_q}{R^2(Q-R)^2(Q-P)^2}- \frac{P^2(1-r_n/q_n)}{R^2(Q-R)^4(Q-P)^2}\nn
&&\hspace{8em}-\frac{P^2(1-p_n/q_n)}{R^2(Q-R)^2(Q-P)^4} \bigg)\,P_T(R)P_T(Q-R)\nn
&&\hspace{8em}+\frac{2P^2}{R^4(Q-R)^2(R-P)^2}P_T(Q)P_T(Q-R) \nn
&&\hspace{8em}+ \frac{P^2}{(Q-P)^4(Q-R)^2(R-P)^2}P_T(R)P_T(Q-R) \Bigg]_Q\nn
&+&\int_{\mathbf{q}\,\mathbf{r}}\frac{\nB(q)}{q} \frac{\nB(r)}{r}\,\Bigg[
2\bigg( \frac{P^2\,f_q}{(Q-R)^2(Q-P)^2}- \frac{P^2(1-r_n/q_n)}{(Q-R)^4(Q-P)^2}\nn
&&\hspace{8em}-\frac{P^2(1-p_n/q_n)}{(Q-R)^2(Q-P)^4} \bigg)\,P_T(R)P_T(Q-R) \nn
&&\hspace{8em}+\frac{2P^2}{(Q+R-P)^2(Q-P)^4}P_T(R)P_T(Q+R)\nn
&&\hspace{8em} + \frac{P^2}{(R-Q-P)^2(Q-R)^4}P_T(Q)P_T(R)\Bigg]_{Q,R}\,.
\ea
Here, the function $f_q$ is associated with the appearance of squared propagators and reads
\begin{equation}
    f_q \equiv \frac{1}{2q^2}+\frac{1+\nB(q)}{2qT},\qquad \textrm{with} \qquad
    -\frac{1}{2q}\frac{\dd}{\dd q}\left( \frac{\nB(q)}{q}\right) = f_q \frac{\nB(q)}{q} \,,
\end{equation}
while the angular brackets stand for (cf.~Ref.~\cite{Laine:2010tc}),
\begin{align}
    [\ldots]_Q & \equiv \frac{1}{2} \sum_{q_n=\pm iq} \{\ldots\}\,,
    &[\ldots]_{Q,R} \equiv \frac{1}{4}\sum_{q_n=\pm iq}\sum_{r_n=\pm ir} \{ \ldots\}\,.
\end{align}
As usual, we have here defined $\nB(q) \equiv (e^{\beta q}-1)^{-1}$ and let $\int_{Q}$ and $\int_{\mathbf{q}}$ stand for $D$ and $D-1$ dimensional integrals, respectively.

The parts of Eq.~(\ref{Ih7c}) proportional to 0, 1 and 2 Bose distribution functions $\nB$ (not counting those inside $f_q$) are henceforth referred to as the 0-, 1- and 2-cut contributions, and are dealt with using different methods. Of the three, the 0-cut piece corresponds to the vacuum ($T=0$) correlator, and can be handled using standard integration-by-parts identities and integral tables, conveniently collected e.g.~in the TARCER Mathematica package \cite{Mertig:1998vk}. This calculation can thus be performed in a fully automated way.

The 1-cut part, on the other hand, involves a three-dimensional thermal integral, in which the integrand is a one-loop vacuum amplitude. A crucial simplification in the evaluation of its UV limit, which we are presently interested in, is that we may perform an expansion in positive powers of the momentum $Q$,
\ba
\label{exp}
\left[\frac{1}{(Q-R)^2}\right]_Q=\left[\frac{1}{R^2}+\frac{2Q\cdot R}{R^4}+\frac{4(Q\cdot R)^2}{R^6}+\cdots\right]_Q\,,
\ea
and similarly with $R\leftrightarrow P$. This relies on the fact that $Q$ is on shell inside the square brackets and the magnitude of the corresponding three-momentum is cut off by the Bose distribution function. The $R$ integrals can then be performed using standard methods, taking advantage of identities such as
\ba
P_T(Q-R)&=&P_T(Q)+P_T(R)-2Q_{\mu}R_{\nu}P_{\mu\nu}^T\,,\nn
\big[(Q\cdot P)^2\big]_Q&=&q^2(\frac{p^2}{D-1}-p_n^2)\,,\qquad (\mathbf{q\cdot p})^2=\frac{q^2p^2}{D-1}\,,
\ea
where we have made use of rotational symmetry.

In the 2-cut part of the expression (\ref{Ih7c}), one may similarly expand the propagators $1/(Q-P)^2$ and $1/(R-P)^2$ in inverse powers of $P^2$. Defining $z\equiv\mathbf{q\cdot r}/(qr)$, $q_n\equiv\sigma\, iq$, $r_n\equiv\rho\, ir$, with $\sigma,\rho=\pm$, we obtain
\ba
\frac{1}{(Q-R)^2}=\frac{1}{2qr(\rho\sigma-z)}
\ea
and subsequently end up dealing with the sums like
\ba
\frac{1}{4}\sum_{\rho=\pm}\sum_{\sigma=\pm}\frac{1}{\rho\sigma-z}&=&\frac{z}{1-z^2}\,,\\
\frac{1}{4}\sum_{\rho=\pm}\sum_{\sigma=\pm}\frac{\rho\sigma}{\rho\sigma-z}&=&\frac{1}{1-z^2}\,,\\
\frac{1}{4}\sum_{\rho=\pm}\sum_{\sigma=\pm}\frac{\rho}{(\rho\sigma-z)^n}&=&\frac{1}{4}\sum_{\rho=\pm}\sum_{\sigma=\pm}\frac{\sigma}{(\rho\sigma-z)^n}=0\,,
\ea
and their derivatives. The remaining task is then to perform the angular integrations. The terms odd
in $\mathbf{q}$ and $\mathbf{r}$ vanish due to antisymmetry, and fixing the
direction of $\mathbf{r}$, we can write the remaining ones in terms of
$z$-averages, \`a la
\ba
\left<\frac{\mathbf{\hat q}\cdot\mathbf{\hat r}\; \mathbf{\hat q}}{1-z^2}\right>_\mathbf{\hat q}&=&\mathbf{\hat r}\left<\frac{z^2}{1-z^2}\right>_z\;=\;\frac{\mathbf{\hat r}}{D-4}\left<1\right>\,,
\ea
where we have made use of rotational invariance and the dimensionally regularized angular integration measure. The other $z$-averages encountered in the calculation are
\ba
\left<\frac{1}{1-z^2}\right>_z &=& \frac{D-3}{D-4}\left<1\right>\,,\\
\left<\frac{1}{(1-z^2)^2}\right>_z &=& \frac{(D-3)(D-5)}{(D-4)(D-6)}\left<1\right>\,,
\ea
while the averages over the angle between $\mathbf{r}$ and $\mathbf{p}$ we need read
\ba
\left<(\mathbf{\hat r}\cdot\mathbf{\hat p})^2\right>_\mathbf{\hat r}=\frac{1}{D-1}\left<1\right>\,,\hspace{2em} \left<(\mathbf{\hat r}\cdot\mathbf{\hat p})^4\right>_\mathbf{\hat r}=\frac{3}{D^2-1}\left<1\right>\,.
\ea

Finally, for clarity of presentation, we wish to remove all $\nB^2(q)$ terms from the result, which is most conveniently done by utilizing the integration by parts identity
\ba
\int_{\mathbf{q}} \nB(q) q^{n-1} f_q = \frac{D-3+n}{2}\int_{\mathbf{q}}\nB(q)q^{n-3}\,.
\ea
Setting $D=4-2\epsilon$, this leads us to the final result for the UV limit of the sum-integral
\ba
\It{12101}{101} & = & -\frac{2^{-12+ 6\epsilon} \pi ^{ -\frac{5}{2}+2 \epsilon} \left(8 \epsilon ^3-24 \epsilon ^2+8 \epsilon +13\right)  \csc (2 \pi  \epsilon ) \Gamma^2(2-\epsilon)}{(1+\epsilon) \Gamma \left(\frac{7}{2}-\epsilon \right) \Gamma \left(5-3\epsilon\right)}\(\fr{\Lambda^{2}}{P^2}\)^{2\epsilon}P^4\nn
& + & \int_{\mathbf{q}} \frac{\nB(q)}{q} \frac{4^{-3+\epsilon } \pi ^{-1+\epsilon } \csc (\pi  \epsilon ) \Gamma (3-\epsilon )}{(3-2\epsilon ) (5-2\epsilon ) \Gamma (2-2 \epsilon )}\(\fr{\Lambda^{2}}{P^2}\)^\epsilon\nn
&\times&\Bigg[P^2 -\frac{4(8 \epsilon ^4-48 \epsilon ^3+103 \epsilon ^2-95 \epsilon +31)}{ (2-\epsilon ) (3-2 \epsilon) }q^2 +2 \epsilon  (1+\epsilon ) q^2 \fr{1}{P^2} \left(\frac{p^2}{3-2\epsilon}-p_n^2\right)\Bigg] \nn
& + & \int_{\mathbf{q}\,\mathbf{r}}\frac{\nB(q)}{q} \frac{\nB(r)}{r} \frac{(1- \epsilon ) (1-2 \epsilon ) (2-\epsilon )  }{ (1+ \epsilon ) (3-2\epsilon ) (5-2\epsilon )} \left[\frac{3  r^2}{q^2} + \frac{2 \epsilon ^3-5 \epsilon ^2+\epsilon -1}{2 \epsilon  (1-2 \epsilon ) (2-\epsilon ) }\right]\nn
&=& -\frac{13}{1440}  \left(\frac{1}{\epsilon}+\frac{5407}{780}\right)\left(\frac{\bar{\Lambda}^2}{P^2}\right)^{2\epsilon}\frac{P^4}{(4\pi )^4}\nn
& + &\frac{1}{15(4\pi )^2}\int_{\mathbf{q}} \frac{\nB(q)}{q} \Bigg[ \frac{3 P^2-62 q^2}{6 \epsilon }+ \left(\frac{ P^2}{2}-\frac{31 q^2}{3}\right)\ln\frac{\bar{\Lambda}^2}{P^2}
+\frac{q^2}{P^2} \left(\frac{p^2}{3}-p_n^2\right)\nn
&+&\frac{47 P^2}{60}+\frac{154 q^2}{45} \Bigg]
 -  \int_{\mathbf{q}\,\mathbf{r}}\frac{\nB(q)}{q} \frac{\nB(r)}{r} \Bigg[\frac{1}{30 \epsilon}- \left(\frac{29}{450}+\frac{2 r^2}{5 q^2}\right) \Bigg]+{\mathcal O}(\epsilon)\,,
\ea
where we have in the last stage performed an expansion in $\epsilon$ and switched to the $\msbar$ scheme. Results for the other sum-integrals, as well as for the full correlator of Eq.~(\ref{Gmasters}), can be obtained in a fully analogous way.

\section{The zero momentum limit of the correlator}

In this Appendix, we determine a perturbative result for the quantity $\tilde G_\eta(0)$, needed in the discussion of Sec.~\ref{4.3}. In the zero momentum limit, the expression (\ref{Gmasters}) consists of vacuum sum-integrals only (recall the definitions of \eqs\nr{eq:defJ} and \nr{eq:defI}) and collapses to
\ba
\frac{\tilde G_\eta(0)}{4 d_A c_\eta^2\Lambda^{2\epsilon}} &=&
D(D-3)\Jt{10}{1}-(D-2)(D-3)\Jt{20}{2} \label{Geta0}\\
&+& \gB^2 \Nc \biggl\{\fr{D(D-2)(D-3)}{2}\(\It{11000}{000}-2\It{21000}{100}\)+2(D-2)^2(D-3)\It{31000}{200} \nn
&-&D(D-3)\It{11100}{100}-\fr{D(D-2)}{2}\(\It{21100}{200}+2\It{21100}{020}
-2\It{11200}{110}-4\It{21100}{110}\)\biggr\} \,.  \nonumber
\ea
Using rotational invariance as well as symmetrizations,
the integrals involving the transverse projection operator
(i.e.~those having non-zero upper indices) can furthermore be related to the more elementary quantities
\ba
I_a^b &=& \Tint{Q}\fr{[q_0]^b}{[Q^2]^a}\,,\quad
S_{abc}^{de}\;=\;\Tint{Q,R}\fr{[q_0]^d [r_0]^e}{[Q^2]^a[R^2]^b[(Q-R)^2]^c} \,,
\ea
via the explicit relations
\ba
\Jt{10}{1}&=&-\fr{D-2}{D-1}I_1^2 \,,\\
\Jt{20}{2}&=&-\fr{D(D-2)}{D^2-1}\Big[2I_1^2-I_2^4\Big] \,,\\
\It{11000}{000}&=&\(I_1^0\)^2 \,,\\
\It{21000}{100}&=&\fr{D-2}{D-1}\Big[I_1^0-I_2^2\Big]I_1^0 \,,\\
\It{31000}{200}&=&\fr{D(D-2)}{D^2-1}\Big[I_1^0-2I_2^2+I_3^4\Big]I_1^0 \,,\\
\It{11100}{100}&=&\fr{D-2}{D-1}\bigg[\(I_1^0\)^2-S_{111}^{20}\bigg] \,,\\
\It{21100}{200}&=&\fr{D(D-2)}{D^2-1}\bigg[\(I_1^0\)^2-2S_{111}^{20}+S_{211}^{40}\bigg] \,,\\
\It{21100}{020}&=&\fr{D(D-2)}{D^2-1}\bigg[\(I_1^0\)^2-2\(I_1^0 I_2^2 + I_1^2 I_2^0\)+S_{211}^{04}\bigg] \,,\\
\It{11200}{110}&=&-\fr{D^2-2D-2}{D^2-1}\bigg[2I_1^2 I_2^0-S_{112}^{22}\bigg]\nn
&-&\fr{1}{2(D^2-1)}\bigg[\(I_1^0\)^2+8I_1^2 I_2^0 -2S_{111}^{20}-4S_{112}^{22}\bigg] \,,\\
\It{21100}{110}&=&\fr{D^2-2D-2}{D^2-1}\bigg[I_1^0\(I_1^0-I_2^2\)-S_{111}^{20}+S_{211}^{22}\bigg]\nn
&+&\fr{1}{2(D^2-1)}\bigg[I_1^0\(3I_1^0-4I_2^2\)-2S_{111}^{20}+4S_{211}^{22}\bigg] \,.
\ea
Using integration-by-parts (IBP) relations on the spatial momentum integrations in a systematic way \cite{Laine:2005ai,Schroder:2008ex}, all two-loop vacuum sum-integrals $S_{abc}^{de}$ reduce to products of one-loop ones $I_a^b$, which by the same strategy (or by direct recursion $I_{a+1}^{b+2}=\frac{2a+1-D}{2a}\,I_a^b$ following from the known analytic solution) can furthermore be related to a few one-loop master sum-integrals. Letting finally $D\rightarrow4$ and plugging the result into Eq.~(\ref{Geta0}), we obtain
\ba
\frac{\tilde G_\eta(0)}{4 c_\eta^2} &=& \fr{8d_A g^2 \Nc}{3}\int_{\mathbf{q}\,\mathbf{r}}\frac{\nB(q)}{q} \frac{\nB(r)}{r} + {\mathcal O}(g^4)
\,=\,\fr{4}{3g^2b_0}\(e-3p\)(T) + {\mathcal O}(g^4) \,.  \label{zeromom}
\ea



\end{document}